\documentclass[hidelinks,man]{apa7} 

\usepackage[style=apa,sortcites=true,sorting=nyt,backend=biber]{biblatex}
\DeclareLanguageMapping{american}{american-apa}

\usepackage[american]{babel} 

\usepackage{csquotes}  

\usepackage{times} 

\usepackage{comment}

\usepackage{amsmath}

\usepackage{amssymb}

\usepackage[symbol]{footmisc} 

\usepackage{tabularx}

\usepackage{listings}  
\definecolor{codegreen}{rgb}{0,0.6,0}
\definecolor{codegray}{rgb}{0.5,0.5,0.5}
\definecolor{codepurple}{rgb}{0.58,0,0.82}
\definecolor{backcolour}{rgb}{0.95,0.95,0.92}
\definecolor{royalblue}{RGB}{65,105,225}
\lstdefinestyle{mystyle}{
    backgroundcolor=\color{backcolour},   
    commentstyle=\color{codegreen},
    keywordstyle=\color{royalblue},
    numberstyle=\tiny\color{codegray},
    stringstyle=\color{codepurple},
    basicstyle=\ttfamily\footnotesize,
    breakatwhitespace=false,         
    breaklines=true,                 
    captionpos=b,                    
    keepspaces=true,                 
    numbers=left,                    
    numbersep=5pt,                  
    showspaces=false,                
    showstringspaces=false,
    showtabs=false,                  
    tabsize=2
}
\title{Evaluating Four Methods for Detecting Differential Item Functioning in Large-Scale Assessments with More Than Two Groups} 
\shorttitle{Evaluating Four Methods}

\author{Dandan Chen Kaptur\textsuperscript{1}\footnote{Corresponding author. Email:\href{mailto:danielle.chen@pearson.com}{\nolinkurl{danielle.chen@pearson.com}}.}, Jinming Zhang\textsuperscript{2}}
\affiliation{\vspace{0.5cm}
\textsuperscript{1} Pearson\\
\textsuperscript{2} University of Illinois Urbana-Champaign}

\abstract{This study evaluated four multi-group differential item functioning (DIF) methods (the root mean square deviation approach, Wald-1, generalized logistic regression procedure, and generalized Mantel-Haenszel method) via Monte Carlo simulation of controlled testing conditions. These conditions varied in the number of groups, the ability and sample size of the DIF-contaminated group, the parameter associated with DIF, and the proportion of DIF items. When comparing Type-I error rates and powers of the methods, we showed that the RMSD approach yielded the best Type-I error rates when it was used with model-predicted cutoff values. Also, this approach was found to be overly conservative when used with the commonly used cutoff value of 0.1. Implications for future research for educational researchers and practitioners were discussed.}

\keywords{differential item functioning, DIF, measurement equivalence, measurement invariance, multi-group DIF, item response theory, RMSD}

\begin{document}

\maketitle

\section{Introduction}

Differential item functioning (DIF) refers to systematic differences in item responses among examinee groups with similar ability levels but from different backgrounds \parencite{american_educational_research_association_standards_2014}. Traditionally, DIF analysis has focused on pairwise comparisons between groups. When more than two groups are involved, one common practice is to use traditional DIF methods designed to analyze two groups at a time \parencite[e.g.,][]{stark_detecting_2006}. This approach can be computationally demanding, inflate Type-I error rates, reduce power, and obscure important patterns of DIF across more than two groups \parencite{penfield_assessing_2001,langer_reexamination_2008}. Additionally, it requires multiple-comparison tests along with the Bonferroni adjustment, which can be conservative and inefficient \parencite{magis_generalized_2011}.

To address these limitations, researchers have developed various \emph{multi-group} DIF detection methods for analyzing more than two groups. One such method, the root mean square deviation (RMSD), has gained traction in large-scale assessments (LSAs), which involve more than two educational systems, ethnicities, languages, cultures, and geographic areas. RMSD's early applications appeared in \textcite{davey_realistic_1997} and \textcite{zhang_chapter_2001}. Today, RMSD is commonly used in LSAs like PISA, PIAAC, TIMSS, and PIRLS \parencite{oecd_pisa_2019,oecd_technical_2019,martin_methods_2020,martin_methods_2017}. Despite its widespread use, the application of a fixed RMSD cutoff lacks empirical grounding. For example, TIMSS employs a cutoff of 0.1 \parencite{martin_methods_2020}, while PISA uses cutoffs of 0.15 and 0.12 \parencite{von_davier_evaluating_2019}, all without justification. While \textcite{tijmstra_sensitivity_2020} found that no universal RMSD cutoff exists due to RMSD's inherent limitations,
RMSD continues to be used with a fixed cutoff \parencite[e.g.,][]{fahrmann_practical_2022,joo_evaluating_2021,buchholz_comparing_2019,kohler_bias-corrected_2020}, and new approaches are emerging based on RMSD using a fixed RMSD cutoff \parencite[e.g.,][]{silva_diaz_performance_2022-1,kohler_semiparametric_2021}.
\\
Although numerous multi-group DIF methods exist, a thorough evaluation comparing their performance has yet to be conducted. Other multi-group DIF methods include the generalized Mantel-Haenszel method \parencite[GMH;][]{somes_generalized_1986}, the generalized logistic regression procedure \parencite[GLR;][]{magis_test-length_2011}, and the improved Wald test \parencite{cai_sem_2008} with Wald-1 \parencite{cai_irtpro_2011} linking algorithm (hereafter referred to as \enquote{Wald-1}). While some studies have attempted to compare multi-group DIF methods \parencite[e.g.,][]{kohler_dif_2024,woods_langer-improved_2013,penfield_assessing_2001,chun_mimic_2016,magis_generalized_2011,chen_performance_2020}, none of them has simultaneously scrutinized the methods mentioned here.
\\
The objective of our study is twofold: We aim to evaluate the performance of commonly used multi-group DIF methods, and to inform the selection of appropriate DIF methods for LSAs. Our evaluation criteria comprised the Type-I error rate and power, assessed using simulated data for each of the selected methods. The Type-I error rate is defined as the proportion of incorrectly identified DIF items out of the total number of items in a test, and the power refers to the proportion of correctly identified DIF items. Both are averaged across replications for each simulated condition.
\\
We focus on frequentist, unidimensional methods within the conventional DIF framework that analyze observed groups without covariates or random effects. They include the following: RMSD, Wald-1, GLR, and GMH. These methods were established for multi-group DIF analyses and have not been compared with each other. 
We exclude mixed-measurement item response theory (IRT) approaches \parencite[e.g.,][]{tay_using_2011} and multi-group latent class analysis \parencite{clogg_simultaneous_1985} as they examine measurement invariance across unobserved classes. We do not consider 
latent factor modeling approaches like multiple-group confirmatory factor analysis \parencite[MGCFA;][]{joreskog_general_1969} and the multiple-indicator multiple-cause method \parencite[MIMIC;][]{joreskog_estimation_1975}, because they analyze latent factors \parencite{chun_mimic_2016,van_de_vijver_invariance_2019}.
Also, we do not involve Bayesian methods such as the Bayesian factor analysis alignment approach \parencite{asparouhov_multiple-group_2014} 
and random effects methods such as GLMM, because they are relatively novel and less frequently used.

\section{Four Multi-Group DIF Methods}

\subsection{Wald-1}

Wald-1 \parencite{cai_irtpro_2011} examines the homogeneity of item parameters based on item response theory \parencite[IRT;][]{richardson_relation_1936}, applicable to multi-group DIF analysis. It uses a linking algorithm along with the approach in \textcite{kim_detection_1995} that extended Lord's Wald test \parencite{lord_applications_1980} to three- and four-group analyses. 
Previous research suggests that Wald-1 outperforms Wald-2 \parencite{langer_reexamination_2008}, while Wald-2 is recommended for identifying anchor items \parencite{woods_langer-improved_2013}.

Wald-1 involves fitting the three-parameter logistic regression model \parencite[3PL;][]{birnbaum_latent_1968} to each item, expressed as
\begin{equation}
\label{eqn:3pl}
    P(X=1 \mid \theta)=c+(1-c)\dfrac{1}{1+e^{-Da(\theta-b)}},
\end{equation}
where\\
\hangindent=0.85in $X$ is a random variable that denotes the item response ($X$ = 1 for correct, $X$ = 0 for incorrect);\\
\indent $a$ is the item discrimination parameter;\\
\indent $b$ is the item difficulty parameter;\\
\indent $c$ is the item pseudo-guessing parameter (also known as the lower asymptote);\\
\hangindent=0.85in $D$ is the scaling factor in IRT, set to 1 throughout this study, following TIMSS practice.\\
\noindent Two-parameter logistic regression model \parencite[2PL;][]{birnbaum_latent_1968} items can be seen as a special case of 3PL items where the $c$ parameter is fixed at zero. 

Given the conceptualization of uniform and nonuniform DIF \parencite{hambleton_item_1985}, in Wald-1, for testing the nonuniform and uniform DIF separately, the null hypothesis is that the IRT-based parameter $a$ or $b$ remains approximately the same across groups. The alternative hypothesis is about the reverse. The null hypotheses are expressed as
\[
    H_0 
    \left\{ 
    \begin{array}{ll}
        a_0 - a_1 = \cdots = a_0 - a_H = 0. & \text{(for nonuniform DIF)} \\
        b_0 - b_1 = \cdots = b_0 - b_H = 0 \text{, if } a_0 - a_1 = \cdots = a_0 - a_H = 0. & \text{(for uniform DIF)}
    \end{array}
    \right.
\]
where\\
\hangindent=0.85in $a_0$ denotes item discrimination for the reference group, and $a_1$ to $a_H$ for focal groups; \\
\hangindent=0.85in $b_0$ denotes item difficulty for the reference group, and $b_1$ to $b_H$ for focal groups.\\
\noindent The alternative hypotheses are
\[
    H_1 
    \left\{ 
    \begin{array}{ll}
        \text{Not all the (} a_0 - a_H \text{) equal 0.} & \text{(for nonuniform DIF)} \\
        \text{Not all the (} b_0 - b_H \text{) equal 0.} & \text{(for uniform DIF)}
    \end{array}
    \right.
\]

\subsection{RMSD}

Similar to Wald-1, RMSD is an IRT-based method. However, RMSD captures DIF by measuring the deviation between a pseudo-observed item characteristic curve (ICC)\footnote{ICC is also known as the probability to answer an item correctly, or \enquote{proportion correct.} Pseudo-observed ICC is estimated based on pseudocounts representing the relative prior expected probabilities of different possibilities.} based on IRT 
and the predicted ICC for the studied item \parencite{oecd_pisa_2017}. The RMSD statistic is expressed as
\begin{equation}
\label{eqn:chap3_rmsd_1}
RMSD=\sqrt{\int [\pi(\theta)-\hat{\pi}(\theta)]^2 g(\theta) \,d\theta},
\end{equation}
where\\
\indent $\theta$ is examinees' abilities;\\
\hangindent=0.85in $\pi(\theta)$ is the pseudo-observed ICC;\\
\indent $\hat{\pi}(\theta)$ is the estimated model-based ICC;\\
\indent $g(\theta)$ is the the probability density of the ability $\theta$ in the population considered in analysis.\\
\noindent For estimation, (\ref{eqn:chap3_rmsd_1}) is rewritten using the Gauss-Hermite quadrature \parencite{pinheiro_approximations_1995} as
\begin{equation}
\label{eqn:chap3_rmsd_2}
\widehat{RMSD}=
\sqrt{\sum_{q=1}^Q [\pi(\theta_q)-\hat{\pi}(\theta_q)]^2 P(\theta_q)},
\end{equation}
where\\
\hangindent=0.85in $P(\theta_q)$ is the (posterior) probability of the examinees observed with a correct answer at the quadrature point $\theta_q$.

The RMSD statistic in nature is a misfit statistic, making RMSD versatile and applicable to more than two groups. As shown in (\ref{eqn:chap3_rmsd_1}), RMSD inherently measures the degree of deviation between pseudo-observed and estimated item response probabilities. Although the RMSD can take the form similar to a pairwise group comparison statistic, shown in (\ref{eqn:chap3_rmsd_2}), its core utility lies in evaluating model fit -- identifying discrepancies between the pseudo-observed ICCs and the model-predicted ICCs for the studied item.

The choice of the RMSD cutoff is crucial, as it determines whether the RMSD statistic indicates a significant deviation. The item is seen as DIF-free if its RMSD statistic is below this cutoff, or DIF-contaminated if otherwise. \textcite{kohler_bias-corrected_2020} demonstrated that using arbitrary cutoff values, such as 0.1 in TIMSS \parencite{martin_methods_2020} and 0.15 and 0.12 in PISA \textcite{von_davier_evaluating_2019}, leads to unsatisfactory Type-I error rates and statistical power. \textcite{chen_modeling_2023} showed that the optimal RMSD cutoff value can be determined using a quadratic regression model. This model takes into account various factors involved in multi-group DIF analysis, with an emphasis on the number of groups.



\subsection{GLR}

GLR \parencite{magis_test-length_2011} is an extension of the logistic regression procedure \parencite{swaminathan_detecting_1990}. Hypothesis testing in GLR is accomplished through the Wald test \parencite{chen_categorical_2023}, similar to the one used for Wald-1 except that different parameters and contrasts are employed \parencite{magis_generalized_2011}. Unlike Wald-1, GLR uses the examinees' total scores as the matching criterion instead of their latent ability, and it does not depend on an IRT model. Its function for DIF analysis is specified below for each group:
\[
    \mbox{logit}(\pi)=\alpha + \beta S,
\]
where\\
\indent $\pi$ is the probability of answering the item correctly;\\
\indent $\alpha$ is the item intercept parameter;\\
\indent $\beta$ is the item slope parameter;\\
\indent $S$ is an examinee's total score.\\
\noindent It can be rewritten into the function below:
\[
    P(X=1 \mid S) = \dfrac{1}{1+e^{-(\alpha + \beta S)}},
\]
where $\alpha + \beta S$ can be compared to $Da (\theta -b)$ in (\ref{eqn:3pl}) for Wald-1. The slope parameter $\beta$ and the intercept parameter $\alpha$ correspond to $Da$ and -$Da b$ in Wald-1. Despite these similarities, GLR differs from Wald-1 in that it does not account for the $c$ parameter in 3PL items, and its proxy of the ability, the total score $S$, is an observed variable.

\subsection{GMH}

GMH \parencite{somes_generalized_1986} is an extension of the Mantel-Haenszel method \parencite[MH;][]{holland_alternate_1985}, applicable to multi-group DIF analysis. Its test statistic is more generalizable than the MH chi-square statistic, useful when more than two categories exist in the condition or the response for each of the $k$ strata, expressed as
\[
\chi^2 = (\boldsymbol{A}-\boldsymbol{E})^\prime \boldsymbol{V}^{-1} (\boldsymbol{A}-\boldsymbol{E}),
\]
where\\
\hangindent=0.85in $\boldsymbol{A}=\sum_k \boldsymbol{A_k}$, where the count of response $j$ in condition $i$ of stratum $k$ is $\boldsymbol{A_k}=(n_{11k}, \cdots, n_{1(r-1)k}, \cdots, n_{(j-1)1k}, \cdots, n_{(j-1)(r-1)k})_{(r-1)(j-1)}$;\\
\hangindent=0.85in $\boldsymbol{E}=\sum_k \boldsymbol{E_k}$, for which the expected value of $\boldsymbol{A_k}^\prime$ is $\boldsymbol{E_k}^\prime=\boldsymbol{j_k}\bigotimes\boldsymbol{r_k}/{n_{\cdot \cdot k}}$ where $\boldsymbol{j_k}=(n_{1 \cdot k},n_{2 \cdot k}, \cdots, n_{(j-1) \cdot k})_{(j-1)}$;\\
\indent $\boldsymbol{V_k}=(n_{\cdot \cdot k}\text{diag}(\boldsymbol{j_k})-\boldsymbol{j_k}{j_k^\prime})\bigotimes(n_{\cdot \cdot k}\text{diag}(\boldsymbol{r_k})-\boldsymbol{r_k r_k}^\prime)/(n_{\cdot \cdot k}-1)n_{\cdot \cdot k}^2$.\\

Unlike Wald-1 and GLR, which rely on estimated model parameters to compute the test statistic, GMH uses observed response frequencies. Similar to MH, GMH uses examinees’ total scores to proxy the ability in DIF testing. Hence, GMH shares some limitations of MH such as the loss of information due to the discretization of the ability parameter and unrealistic assumptions about equally spaced ability values and a constant common odds ratio $\alpha$ across distinct ability levels \parencite{zwick_review_2012}. Despite these limitations, GMH has demonstrated superior power and adequate control over Type-I error \parencite{penfield_assessing_2001}. \textcite{finch_detection_2016} found that GMH provides an optimal balance of Type-I error control and power among the tested frequentist methods. 

\section{Method}

The current literature suggests that more variation is needed in Monte Carlo simulations to evaluate multi-group DIF methods. Previous studies have limited their scope to three groups for Wald-1 \parencite{woods_langer-improved_2013}, up to six groups for GLR and GMH \parencite{finch_detection_2016}, but up to 50 groups for RMSD \parencite{kohler_dif_2024}. Also, there is a lack of scrutiny when manipulating the groups' ability level, the item discrimination parameter $a$, or the sample sizes of multiple groups, while the existing literature has unveiled the following: RMSD cannot consistently detect DIF associated with low-performing groups \parencite{tijmstra_sensitivity_2020} or the item discrimination parameter \parencite{buchholz_comparing_2019}; GLR's performance can be affected by unequal sample sizes \parencite{finch_detection_2016}.

Our study was intended to address these limitations. We enhanced the scope of existing literature by including up to 15 examinee groups, reflecting the common scenario in LSAs, which often involve 15 or more country-differentiated examinee groups. For a baseline comparison, we also included two-group analyses. We manipulated several understudied factors such as the ability distribution of the DIF-contaminated focal group, the DIF-related item parameter, the sample size of the DIF-contaminated focal group, in addition to commonly studied factors. To maintain clear and interpretable results, we focused on the simplest simulated conditions where only one focal group was affected by DIF.

We set the significance level $\alpha$ at 0.05 and the prior power $(1-\beta)$ at 0.8, and reported the Type-I error rate and power averaged across 100 replications for each simulated condition. This number of replications was chosen given the latest publications in this field \parencite[e.g.,][]{lim_residualbased_2022,kohler_dif_2024} as well as our computation: We used the formula $n=(\frac{Z_{\alpha/2}-Z_{\beta}}{(\mu_1-\mu_0)/\sigma})^2$ \parencite{lomax_introduction_2012,web_interface} where $Z_{\alpha/2} = 1.96$, $Z_{\beta} = -0.84$, $\mu_0 = 0.05$, and $\mu_1$ and $\sigma$ are means and standard deviations of Type-I error rates from an initial exploration with 30 replications across all of our simulated conditions\footnote{The numbers we obtained were averaged at 7, with the 25th percentile at 2 and the 75th percentile at 21.}. Our acceptable range for the Type-I error rate $\alpha$ was 0.01 to 0.09, calculated based on a Bernoulli distribution using the formula $0.05 \pm 1.96\frac{\sqrt{0.05(1-0.05)}}{\sqrt{100}}$, following \textcite{woods_langer-improved_2013}. 

Our simulations were divided into Study 1, where we assessed the methods with DIF-free data, and Study 2, where we examined the methods when one focal group was set to be DIF-contaminated. Specifically, Study 1 varied the number of groups (i.e., 2, 5, 10, 15), the ability level (i.e., low, high), and the sample size (i.e., small, large) of the focal group set to be DIF-contaminated in Study 2. Study 2 manipulated additional factors: the parameter associated with DIF (difference of 0.4 in the $a$ or $b$ parameter) and the proportion of DIF items (around 20\% and 30\%, corresponding to 6 and 9 items). We randomized DIF-contaminated items for each replication to attenuate item-associated effects, following \textcite{woods_langer-improved_2013}. Also, we assigned the DIF-contaminated focal group consistently higher values in either the \emph{a} parameter or the \emph{b} parameter compared to the other groups, following \textcite{penfield_assessing_2001}. Overall, these conditions were determined given the ones that have appeared in the existing literature and our experience with DIF analyses using the real-world LSA data. Consequently, for two variations of the four methods we tested, there were 16 simulated conditions in Study 1 (4 group numbers $\times$ 2 ability levels $\times$ 2 sample sizes) and 64 simulated conditions in Study 2 (16 conditions from Study 1 $\times$ 2 parameters for DIF $\times$ 2 proportions of DIF items).

\subsection{Procedures for data simulation}

The groups in our simulation refer to examinee groups differentiated by national or regional education systems. Our data was simulated to align with actual data from the TIMSS 2019 Mathematics Grade-8 assessment \parencite{fishbein_timss_2021}. TIMSS's Booklet 13 was randomly selected for obtaining the test length, item parameter estimates, ability distributions, and sample sizes for various groups that were used in this simulation. Table \ref{tab:grpdta} documents the mean and standard deviations of the ability $\theta$ distribution and the sample size we used for each group.
For each simulated condition, 100 data sets were generated for replications using values drawn randomly from the ability distribution for each group.
\\
We chose \enquote{Western Cape, RSA} as the reference group because its estimated average $\theta$ was close to 0 and because it was included as a benchmark region for TIMSS 2019. 
We used \enquote{Kuwait,} \enquote{Romania,} \enquote{Morocco,} and \enquote{Australia} to simulate the DIF-contaminated focal group in each scenario: (1) small sample size, low ability ($N$ = 327, $\mu_{\theta}$ = -0.393 for Kuwait); (2) small sample size, high ability ($N$ = 326, $\mu_{\theta}$ = 0.344 for Romania); (3) large sample size, low ability ($N$ = 598, $\mu_{\theta}$ = -0.560 for Morocco); (4) large sample size, high ability ($N$ = 642, $\mu_{\theta}$ = 0.773 for Australia). We randomly selected the DIF-free focal groups from the rest of the groups, ensuring that they were a balanced mix of low- and high-ability groups, each with more than 300 examinees. 
\\
It is worth noting that we chose \enquote{small} and \enquote{large} to label sample sizes here to ensure relevance and applicability to LSAs like TIMSS. While the sample sizes we labeled as \enquote{large} (598 for Morocco or 642 for Australia) may seem modest within the broader context of LSAs, they correspond to the large sample sizes that truly exist in the TIMSS assessment data set. These varying sample sizes are sufficient to reveal whether the tested methods may be influenced by sample size. By defining \enquote{large} and \enquote{small} sample sizes using real-world data, we strike a balance between theoretical exploration and practical relevance.
\\
Table \ref{tab:itempar} documents the IRT model, item parameters and other information about 29 items from the selected booklet. All these item parameters aligned with the appropriate range in testing. All items were utilized as 3PL items since 2PL is a special case of 3PL where the $c$ parameter is fixed at zero. We primarily followed the latest practice to manipulate differences in the IRT item parameters to simulate DIF \parencite[e.g.,][]{lim_residualbased_2022,kohler_dif_2024,rutkowski_assessing_2014}. To allow the comparison with early DIF simulation studies, we also computed the corresponding difference in terms of the ICC area \parencite[]{rudner_approach_1977} and the effect size of DIF \parencite[]{holland_differential_1986,zwick_review_2012}, shown in the last six columns of Table \ref{tab:itempar}. This computation was done when using 41 quadrature points of the ability $\theta$ on the scale [-4, 4] (or \textit{Q41}, which has increments of 0.2), given the existing evidence about different scaling options supporting measurement precision \parencite[e.g.,][]{antal_adaptive_2007,mazumder_numerical_2016}.
The ICC area difference varied from 0.041 to 0.309 for the $a$ parameter and from 0.270 to 0.400 for the $b$ parameter. The effect size $\Delta_{MH}$ ranged from 0.000 to 0.057 for the $a$ parameter and from 0.416 to 2.551 for the $b$ parameter. Manipulating the $a$ parameter resulted in DIF items flagged with an \enquote{A} rating, while changes in the $b$ parameter led to a mix of \enquote{A,} \enquote{B,} and \enquote{C} ratings \parencite{zwick_review_2012}.

\begin{table}
\small
\centering
\begin{threeparttable}
\caption{\label{tab:grpdta}Group-specific statistics used in the data simulation.}
\def\arraystretch{0.7} 
\begin{tabular}{p{0.4\linewidth} >{\raggedleft\arraybackslash}p{0.05\linewidth} >{\raggedleft\arraybackslash}p{0.08\linewidth} >{\raggedleft\arraybackslash}p{0.08\linewidth}}  
\hline
\multicolumn{1}{c}{ } & \multicolumn{1}{c}{ } & \multicolumn{2}{c}{Ability\tnote{ii}} \\
\cmidrule(r){3-4}
Group & \multicolumn{1}{c}{$N$\tnote{i}} & \multicolumn{1}{c}{$\hat{\mu}$} & \multicolumn{1}{c}{$\hat{\sigma}$}\\
\hline
Australia & 642 & 0.773 & 0.826\\

Bahrain & 410 & 0.296 & 1.058\\

Iran & 432 & -0.008 & 0.851\\

Jordan & 518 & -0.316 & 0.728\\

Kuwait & 327 & -0.393 & 0.794\\

Lebanon & 340 & -0.272 & 0.675\\

Morocco & 598 & -0.560 & 0.589\\

Oman & 483 & -0.204 & 0.817\\

New Zealand & 437 & 0.524 & 0.872\\

Romania & 326 & 0.344 & 0.933\\

Saudi Arabia & 405 & -0.409 & 0.718\\

South Africa & 1480 & -0.369 & 0.729\\

Egypt & 521 & -0.240 & 0.828\\

Gauteng, RSA & 409 & -0.230 & 0.708\\

Western Cape, RSA & 374 & -0.119 & 0.841\\
\hline
\end{tabular}
\begin{tablenotes}
        \item[i] Sample size of examinees in the selected booklet data.\\
        \item[ii] Estimates based on the selected booklet data.
    \end{tablenotes}
\end{threeparttable}
\end{table}

\begin{table}
\footnotesize
\centering
\begin{threeparttable}  
\caption{\label{tab:itempar}Item parameters used in the data simulation.}
\def\arraystretch{0.7} 
\begin{tabular}{rlrrrrrrrrrrrr}
\hline
\multicolumn{1}{c}{ } & \multicolumn{1}{c}{ } & \multicolumn{1}{c}{ } & \multicolumn{3}{c}{DIF-Free Groups\tnote{i}} & \multicolumn{2}{c}{DIF Groups\tnote{ii}} & \multicolumn{3}{c}{DIF in a} & \multicolumn{3}{c}{DIF in b} \\
\cmidrule(r){4-6} \cmidrule(r){7-8} \cmidrule(r){9-11} \cmidrule(r){12-14}
No. & Item & Model & $\hat{a}$\tnote{iii} & $\hat{b}$ & $\hat{c}$ & $a$ & $b$ & Area & $\Delta_{MH}$ & Flag & Area & $\Delta_{MH}$ & Flag\\
\hline
1 & MP62001 & 3PL & 1.219 & 1.134 & 0.299 & 1.619 & 1.534 & 0.184 & 0.018 & A & 0.270 & 0.416 & A\\

2 & MP62067 & 3PL & 2.050 & 0.360 & 0.312 & 2.450 & 0.760 & 0.076 & 0.000 & A & 0.275 & 0.522 & A\\

3 & MP62072 & 2PL & 1.682 & 0.275 & 0.000 & 2.082 & 0.675 & 0.158 & 0.002 & A & 0.399 & 1.581 & C\\

4 & MP62120 & 3PL & 2.466 & 0.903 & 0.145 & 2.866 & 1.303 & 0.067 & 0.000 & A & 0.342 & 0.843 & A\\

5 & MP62146 & 3PL & 2.079 & 1.181 & 0.109 & 2.479 & 1.581 & 0.095 & 0.002 & A & 0.355 & 0.906 & A\\

6 & MP62154 & 2PL & 2.050 & 0.278 & 0.000 & 2.450 & 0.678 & 0.110 & 0.001 & A & 0.400 & 1.927 & C\\

7 & MP62192 & 2PL & 1.440 & 1.696 & 0.000 & 1.840 & 2.096 & 0.194 & 0.057 & A & 0.384 & 1.354 & C\\

8 & MP62214 & 2PL & 1.826 & 0.849 & 0.000 & 2.226 & 1.249 & 0.135 & 0.005 & A & 0.398 & 1.716 & C\\

9 & MP62242 & 3PL & 2.014 & 0.524 & 0.189 & 2.414 & 0.924 & 0.092 & 0.000 & A & 0.324 & 0.708 & A\\

10 & MP62250A & 2PL & 1.988 & 0.552 & 0.000 & 2.388 & 0.952 & 0.116 & 0.002 & A & 0.399 & 1.869 & C\\

11 & MP62250B & 2PL & 2.485 & 1.211 & 0.000 & 2.885 & 1.611 & 0.077 & 0.001 & A & 0.399 & 2.336 & C\\

12 & MP62341 & 3PL & 2.356 & 1.873 & 0.225 & 2.756 & 2.273 & 0.065 & 0.002 & A & 0.307 & 0.538 & A\\

13 & MP72005 & 3PL & 1.105 & 0.093 & 0.026 & 1.505 & 0.493 & 0.309 & 0.004 & A & 0.381 & 0.941 & A\\

14 & MP72021 & 2PL & 1.526 & 0.604 & 0.000 & 1.926 & 1.004 & 0.185 & 0.008 & A & 0.397 & 1.435 & C\\

15 & MP72026 & 2PL & 1.131 & 0.990 & 0.000 & 1.531 & 1.390 & 0.297 & 0.055 & A & 0.385 & 1.064 & C\\

16 & MP72041A & 2PL & 1.645 & 0.275 & 0.000 & 2.045 & 0.675 & 0.164 & 0.002 & A & 0.399 & 1.547 & C\\

17 & MP72041B & 2PL & 2.038 & 0.615 & 0.000 & 2.438 & 1.015 & 0.111 & 0.002 & A & 0.399 & 1.916 & C\\

18 & MP72059 & 2PL & 2.159 & 0.978 & 0.000 & 2.559 & 1.378 & 0.100 & 0.002 & A & 0.399 & 2.030 & C\\

19 & MP72080 & 3PL & 2.428 & 1.269 & 0.098 & 2.828 & 1.669 & 0.073 & 0.001 & A & 0.360 & 1.008 & B\\

20 & MP72081 & 2PL & 1.314 & 1.415 & 0.000 & 1.714 & 1.815 & 0.229 & 0.056 & A & 0.385 & 1.235 & C\\

21 & MP72094 & 2PL & 1.970 & 0.238 & 0.000 & 2.370 & 0.638 & 0.118 & 0.001 & A & 0.400 & 1.852 & C\\

22 & MP72120 & 2PL & 1.820 & 1.284 & 0.000 & 2.220 & 1.684 & 0.135 & 0.011 & A & 0.397 & 1.711 & C\\

23 & MP72131 & 2PL & 2.126 & 1.811 & 0.000 & 2.526 & 2.211 & 0.100 & 0.013 & A & 0.395 & 1.998 & C\\

24 & MP72140 & 2PL & 1.327 & 0.761 & 0.000 & 1.727 & 1.161 & 0.234 & 0.021 & A & 0.393 & 1.248 & C\\

25 & MP72147 & 2PL & 2.714 & 1.625 & 0.000 & 3.114 & 2.025 & 0.065 & 0.002 & A & 0.399 & 2.551 & C\\

26 & MP72154 & 3PL & 2.021 & 0.581 & 0.179 & 2.421 & 0.981 & 0.093 & 0.001 & A & 0.328 & 0.728 & A\\

27 & MP72161 & 2PL & 1.911 & 1.149 & 0.000 & 2.311 & 1.549 & 0.124 & 0.007 & A & 0.398 & 1.796 & C\\

28 & MP72192 & 3PL & 2.013 & 0.901 & 0.216 & 2.413 & 1.301 & 0.089 & 0.001 & A & 0.313 & 0.616 & A\\

29 & MP72223 & 3PL & 2.995 & 0.967 & 0.244 & 3.395 & 1.367 & 0.041 & 0.000 & A & 0.302 & 0.604 & A\\
\hline
\end{tabular}
    \begin{tablenotes}
        \item[i] Estimates based on the selected booklet data.\\
        \item[ii] Estimates set to be 0.4 higher than the DIF-free group.\\
        \item[iii] The scaling factor $D$ in IRT was set to 1, following TIMSS practice.
    \end{tablenotes}
\end{threeparttable}
\end{table}

\subsection{Settings for data analysis}

We used three software packages for analyses: the R package {\fontfamily{qcr}\selectfont mirt} \parencite{chalmers_multidimensional_2021} for RMSD, the program {\fontfamily{qcr}\selectfont flexMIRT} \parencite{houts_flexmirt_2020} for Wald-1, and the R package {\fontfamily{qcr}\selectfont difR} \parencite{chalmers_multidimensional_2021} for GLR and GMH. The base R version was R 4.2.1 \parencite{r_core_team_r_2021}. We employed the default options implemented in the software programs to recover the asymptotic covariance matrix.
We used maximum likelihood \parencite[ML;][]{fisher_mathematical_1922} for parameter estimation for RMSD and maximum a posteriori (MAP) for Wald-1, following the common practice for parameter estimation in different methods. We used both the arbitrary cutoff or RMSD (i.e., 0.1 from TIMSS 2019) and model-predicted cutoffs from \textcite{chen_modeling_2023} (i.e., 0.060, 0.070, 0.075, and 0.075 for 2-, 5-, 10- and 15-group analyses, respectively).
\\
Wald-1, GLR, and GMH require anchor items. We utilized the constant anchor item method, analyzing the non-anchor items for DIF, due to its superiority over other methods \parencite[e.g.,][]{wang_effects_2003,wang_effects_2004,kopf_anchor_2015}. We identified anchor items using Wald-2, following \textcite{woods_langer-improved_2013}. If all items were found DIF-free, we used the second half of the test items as anchor items. 
\\
For $p$-value adjustment in GLR and GMH, we used Holm’s adjustment method \parencite{holm_simple_1979}, preferred over the Bonferroni procedure \parencite{simes_improved_1986}. This adjustment is deemed necessary for GLR due to inflated Type-I errors associated with analyzing numerous groups \parencite{magis_generalized_2011}. For GMH, we reported results with and without the $p$-value adjustment to explore the optimal option due to limited research in this area.

\section{Results}

\subsection{Study 1}

\begin{figure}
    \includegraphics[width=\textwidth]{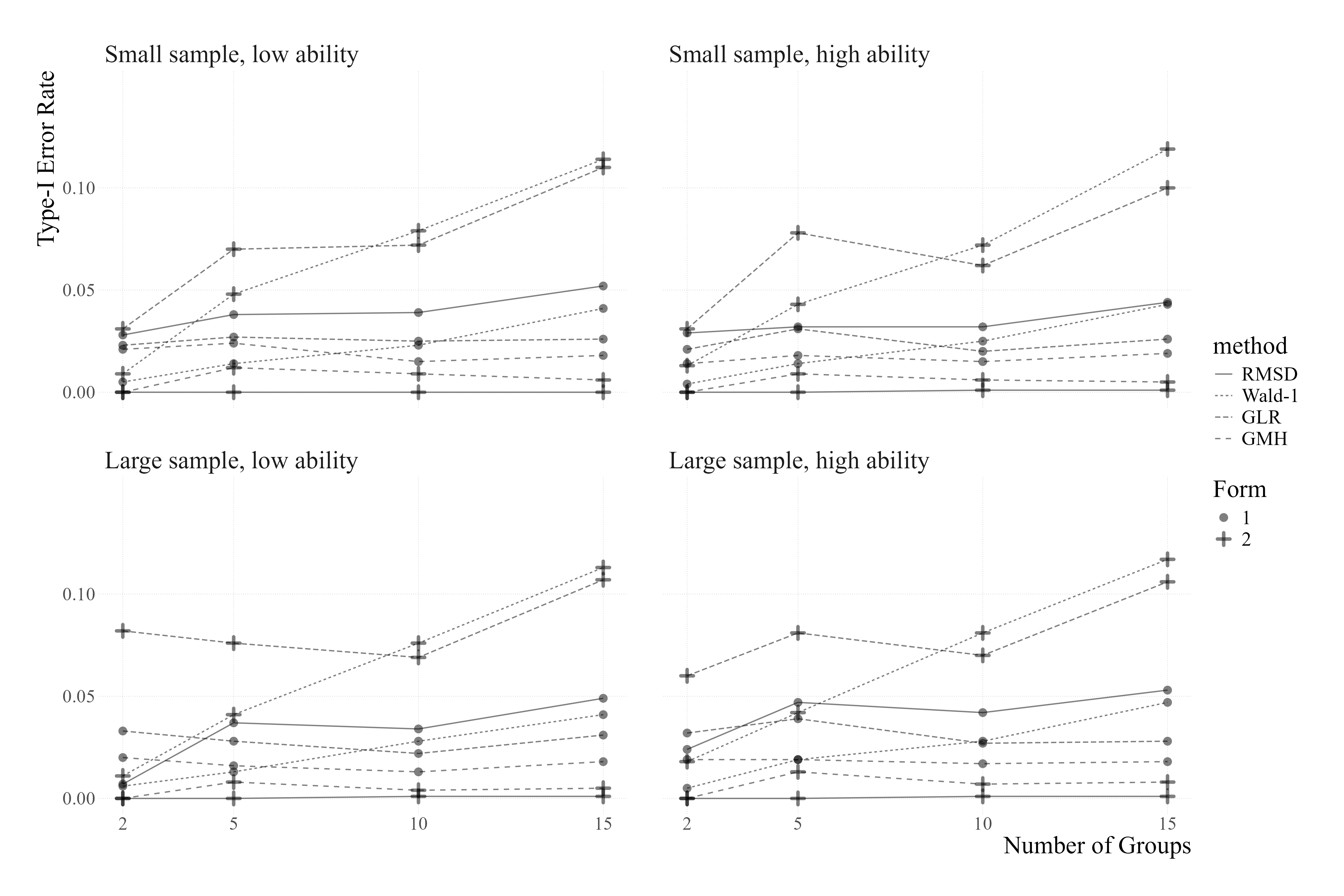}
    \caption{Type-I error rates with the selected methods in DIF detection (using DIF-free data).}
\label{fig:methods_typ1}
\end{figure}

Table \ref{tab:methods_typ1} and Figure \ref{fig:methods_typ1} show the statistics about Type-I error rates in simulated DIF-free conditions. They present error rates across various conditions, for two variations of each method (i.e., Form 1 and Form 2), when varying the DIF-contaminated focal group in the sample size and the ability level and varying the number of groups from two to 15.
\\
The results indicate that RMSD with the arbitrary cutoff exhibited near-zero Type-I error rates. They were overly conservative, meaning RMSD with the arbitrary cutoff is less likely to flag items as having DIF even in cases where some DIF might be present. However, RMSD with model-predicted cutoffs displayed acceptable Type-I error rates, which were particularly noteworthy because they were the closest to the nominal alpha level, indicating that RMSD correctly identified items without DIF most of the time when using with model-predicted cutoffs.
\\
GMH method showed satisfactory performance when not using the $p$-value adjustment, as these rates were within the acceptable range. However, GMH became overly conservative when using the $p$-value adjustment, demonstrated by its excessively low Type-I error rates. This conservative nature with $p$-value adjustment, while reducing false positives, may inadvertently increase the risk of false negatives, making GMH less suitable for scenarios where identifying all potential DIF items is critical.
\\
Wald-1 and GLR were shown to be useful for detecting both uniform and nonuniform DIF. However, they exhibited inflated Type-I error rates when detecting nonuniform DIF. Also, their performance varied greatly with the number of groups involved in the analysis. In particular, their Type-I error rates for nonuniform DIF analyses with 15 groups exceeded the acceptable range [0.01, 0.09]. This suggests that Wald-1 and GLR may be less reliable when dealing with a large number of groups, particularly for nonuniform DIF detection.
\\
In summary, while each method has its strengths and limitations, RMSD with its model-predicted cutoff showed optimal performance in maintaining low Type-I error rates. GMH turned overly conservative with the $p$-value adjustment. Wald-1 and GLR were versatile for DIF detection but struggled with larger group analyses, particularly for nonuniform DIF detection.

\subsection{Study 2}

\begin{figure}
    \includegraphics[width=0.95\textwidth]{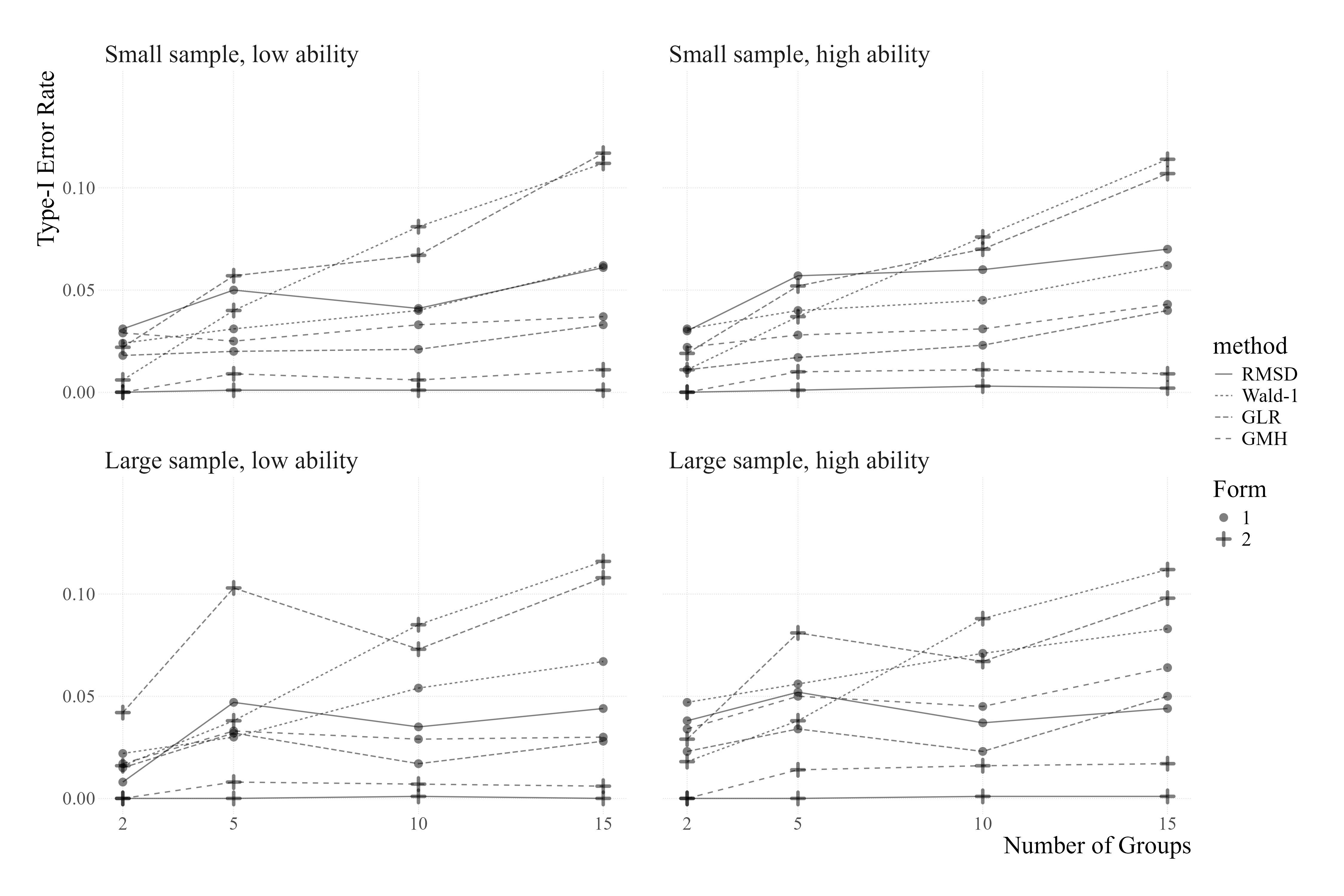}
    \caption{Type-I error rates (with DIF in $b$ for 20\% items).}
    \label{fig:methods_typ1_1}

    \bigskip

    \includegraphics[width=0.95\textwidth]{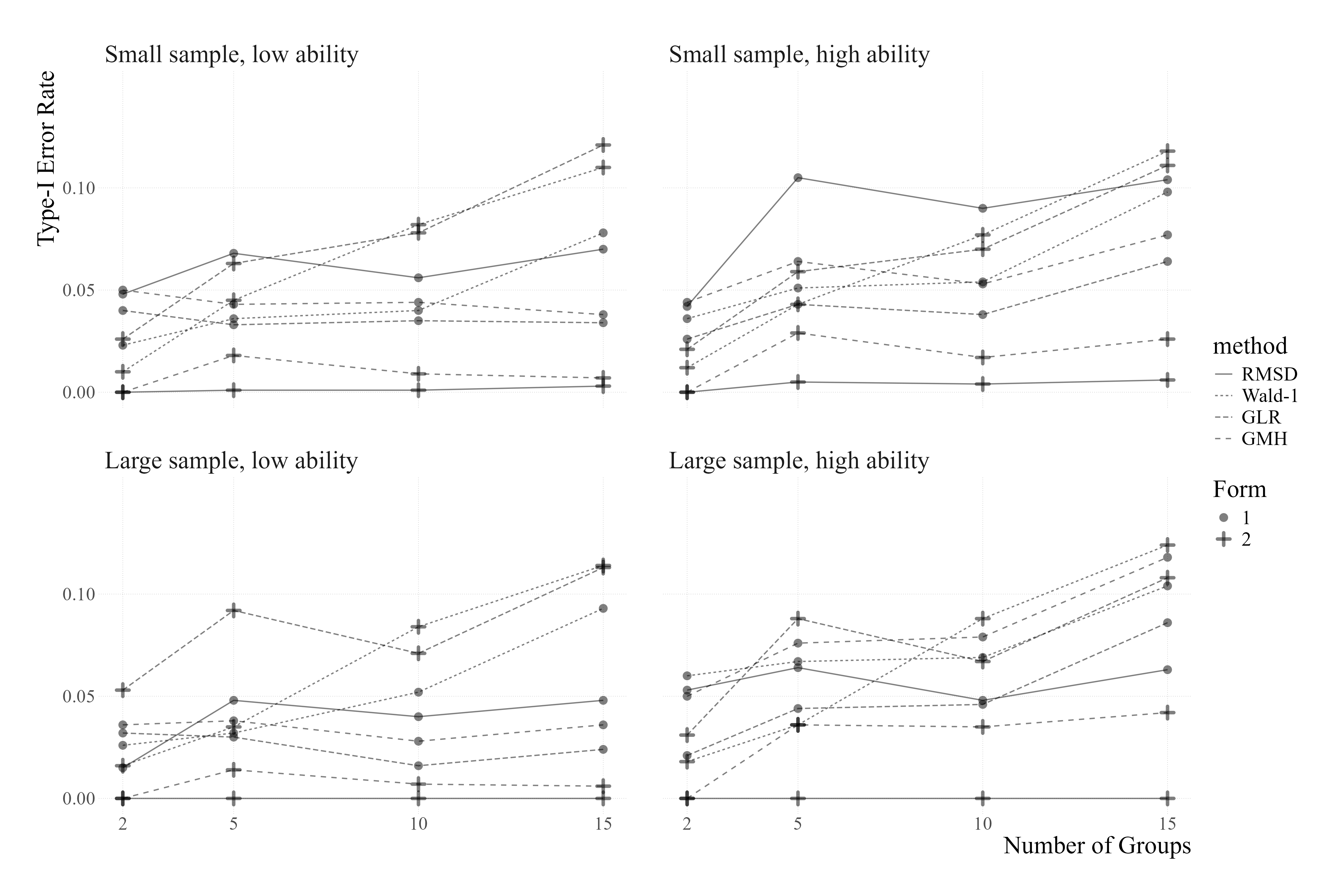}
    \caption{Type-I error rates (with DIF in $b$ for 30\% items).}
\label{fig:methods_typ1_2}
\end{figure}

\begin{figure}
    \includegraphics[width=0.95\textwidth]{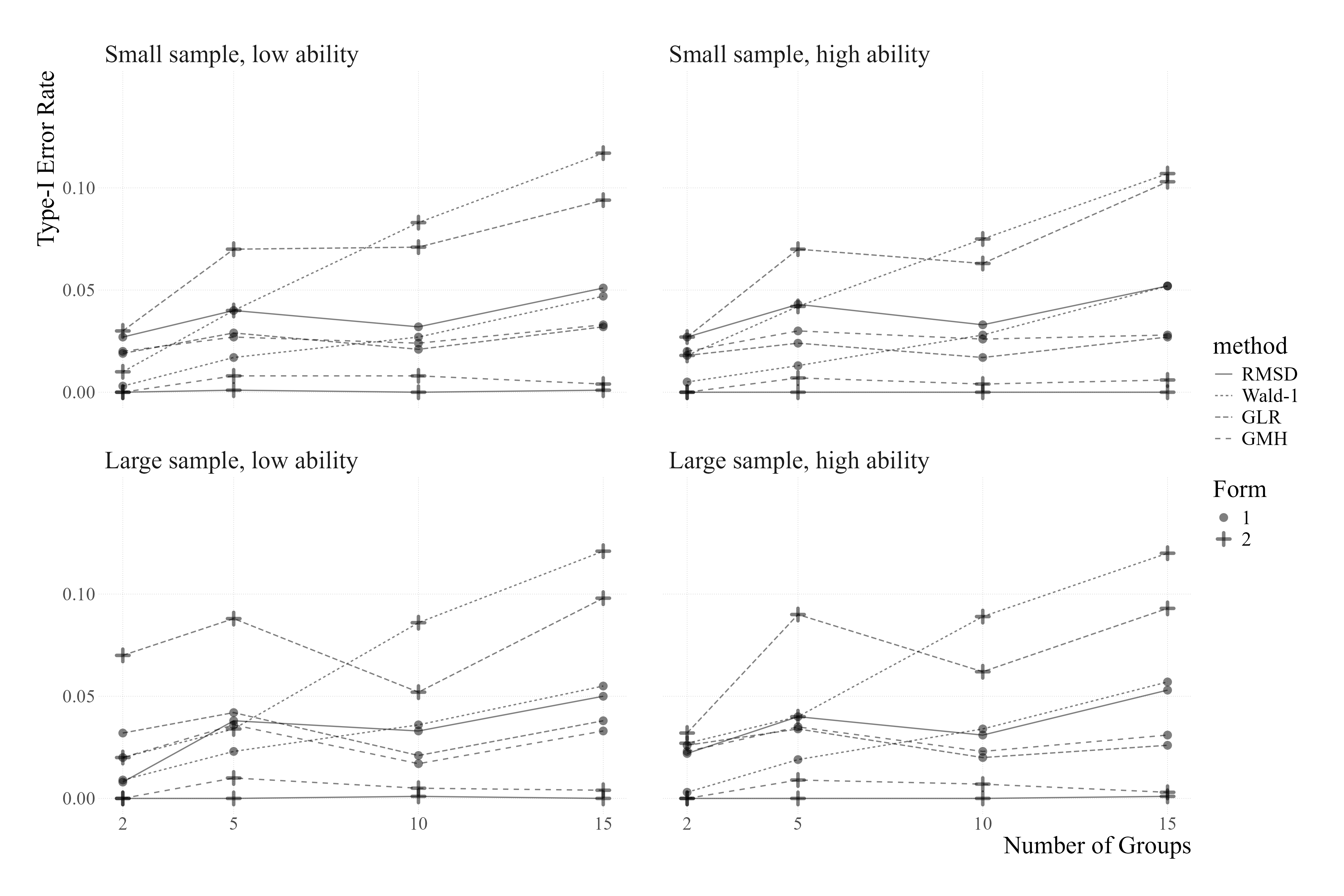}
    \caption{Type-I error rates (with DIF in $a$ for 20\% items).}
    \label{fig:methods_typ1_3}

    \bigskip
    
    \includegraphics[width=0.95\textwidth]{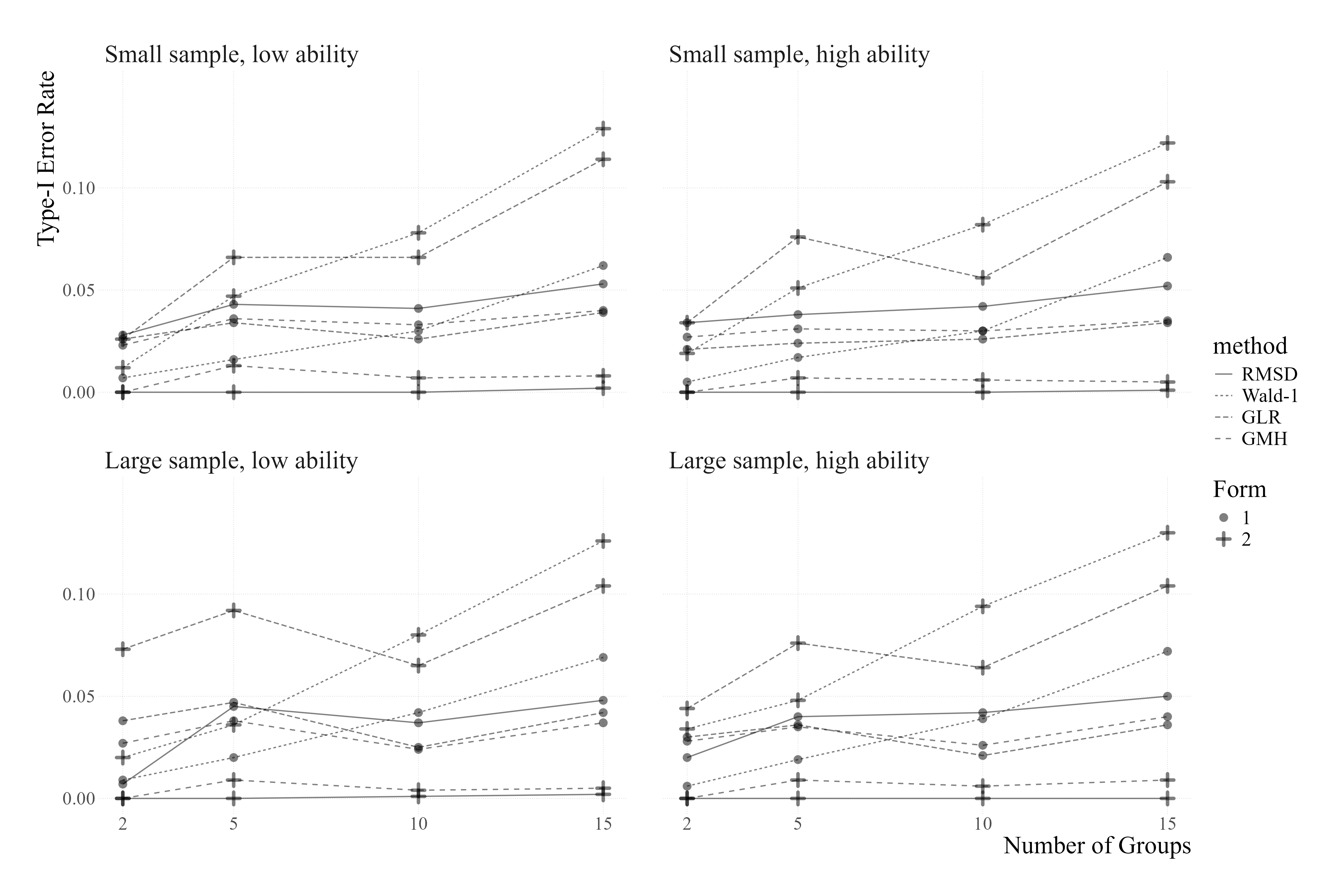}
    \caption{Type-I error rates (with DIF in $a$ for 30\% items).}
\label{fig:methods_typ1_4}
\end{figure}

\begin{figure}
    \includegraphics[width=0.95\textwidth]{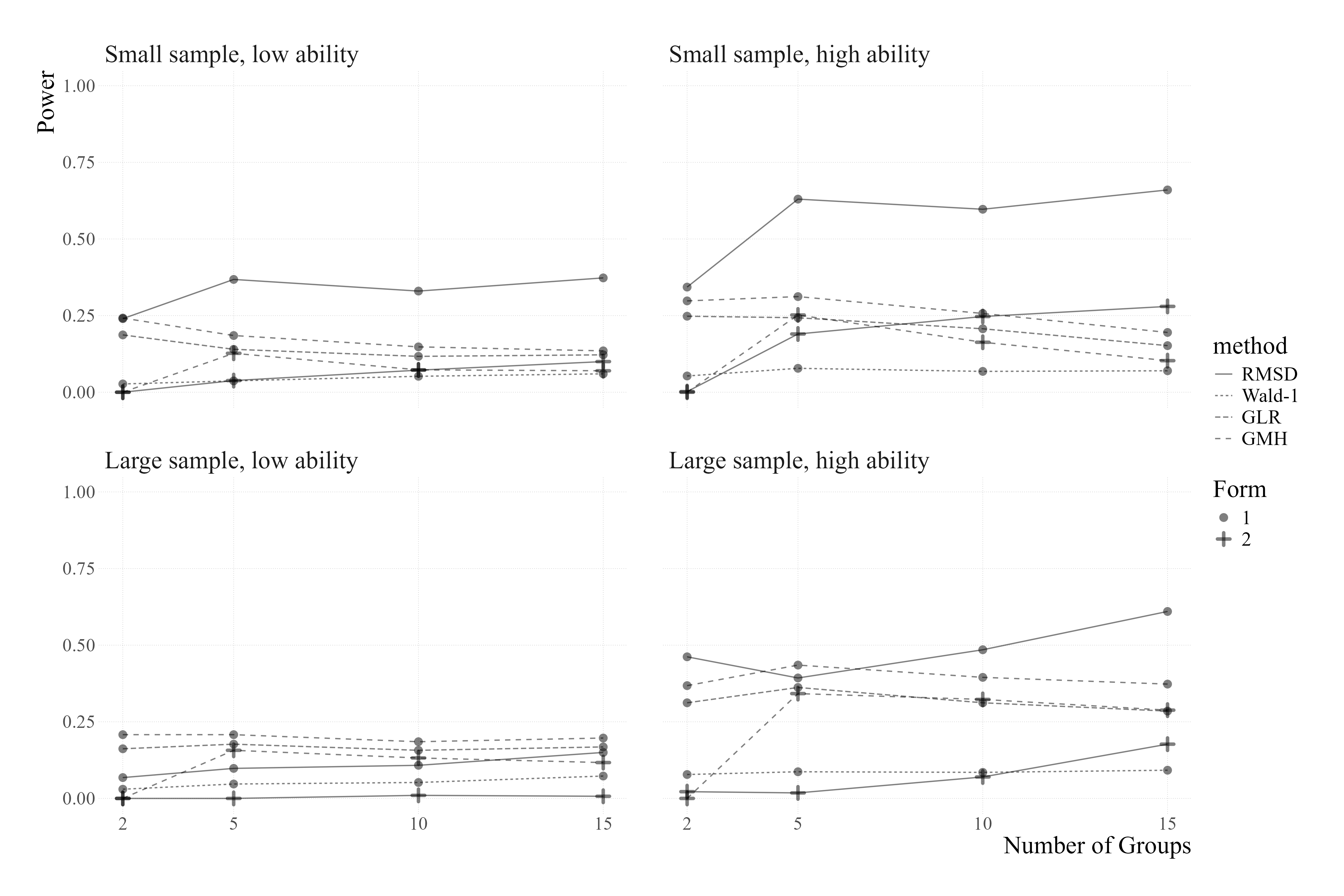}
    \caption{Powers with the selected methods in DIF detection (with DIF in $b$ for 20\% items).}
    \label{fig:methods_power_1}

    \bigskip

    \includegraphics[width=0.95\textwidth]{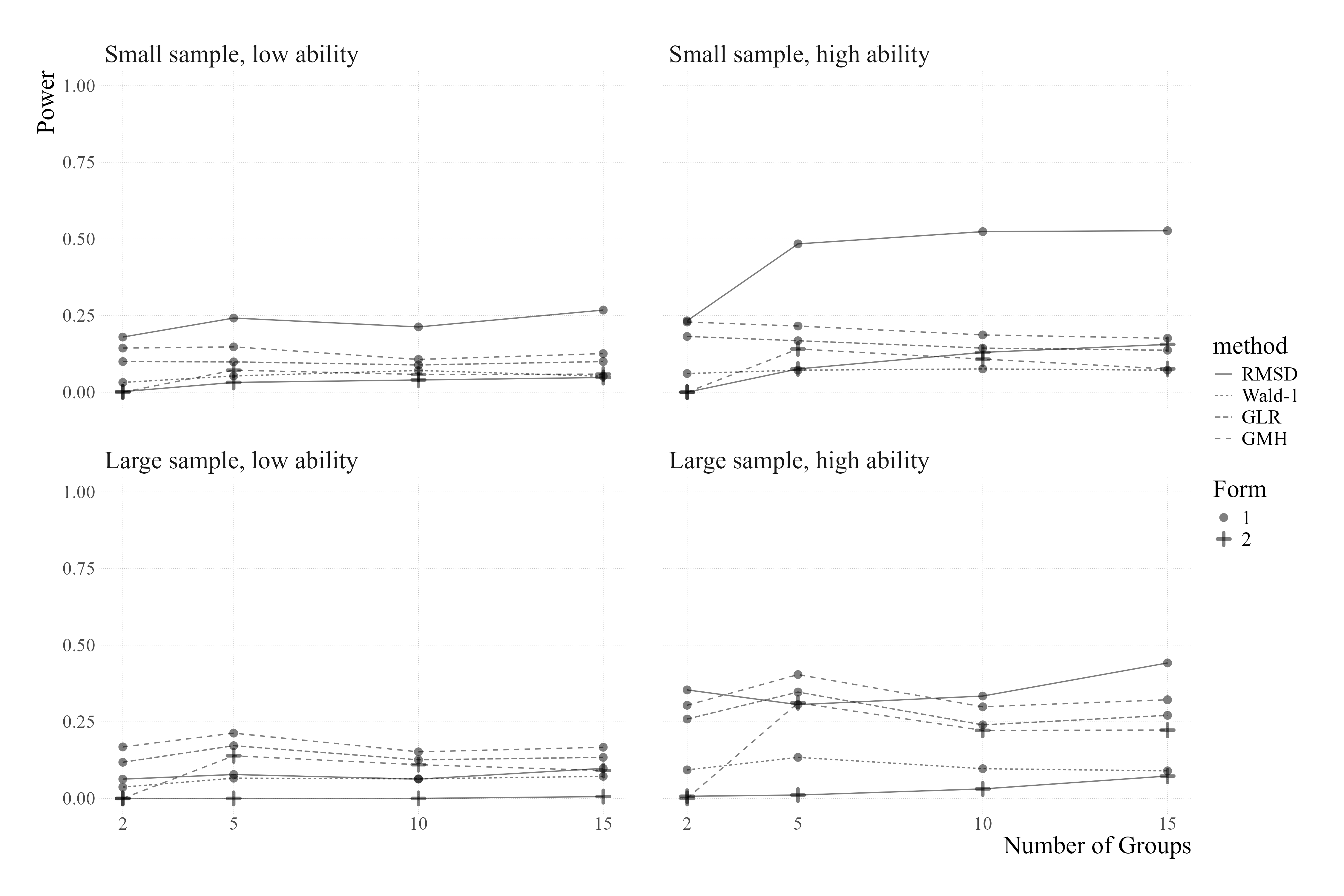}
    \caption{Powers with the selected methods in DIF detection (with DIF in $b$ for 30\% items).}
\label{fig:methods_power_2}
\end{figure}

\begin{figure}
    \includegraphics[width=0.95\textwidth]{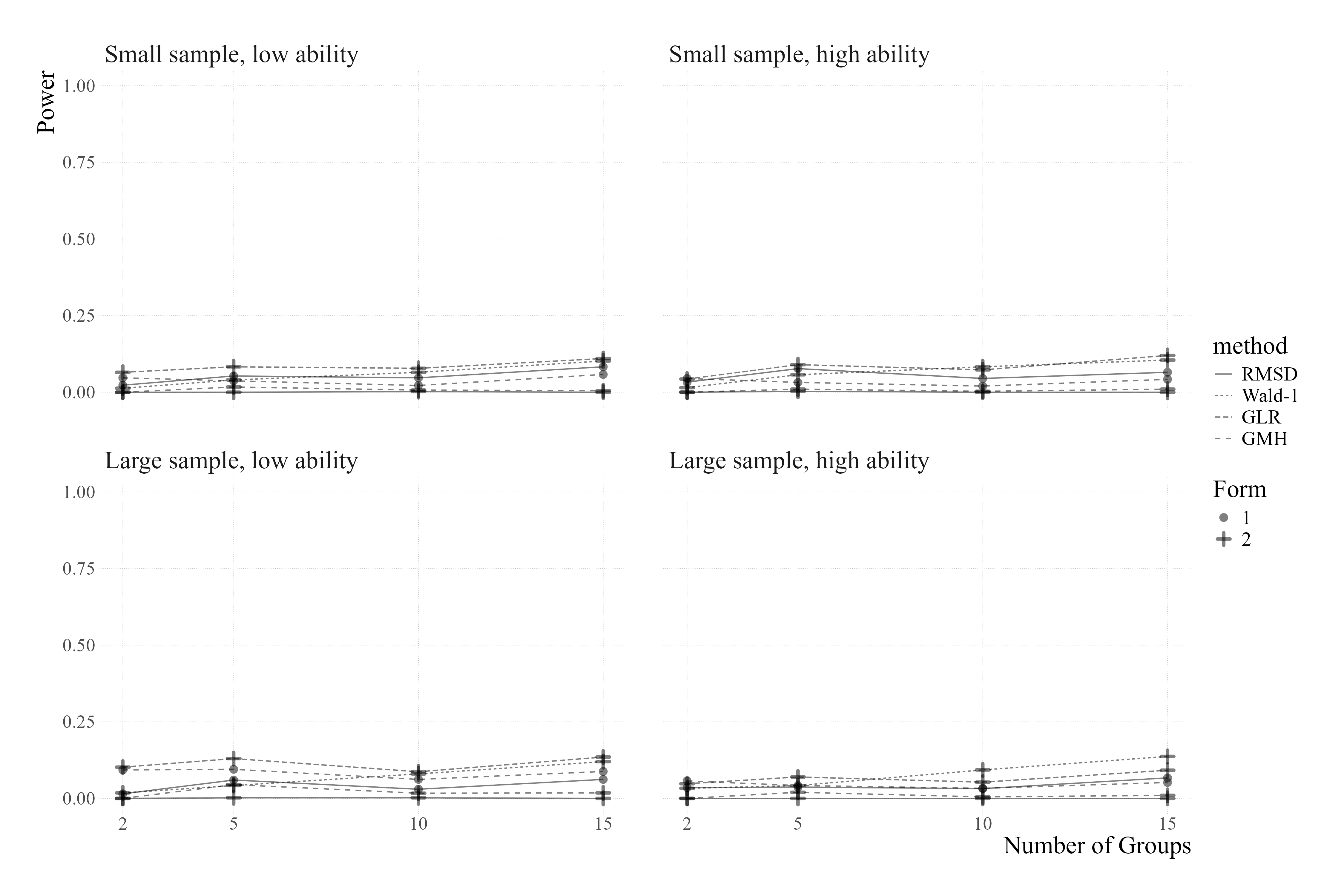}
    \caption{Powers with the selected methods in DIF detection (with DIF in $a$ for 20\% items).}
    \label{fig:methods_power_3}

    \bigskip

    \includegraphics[width=0.95\textwidth]{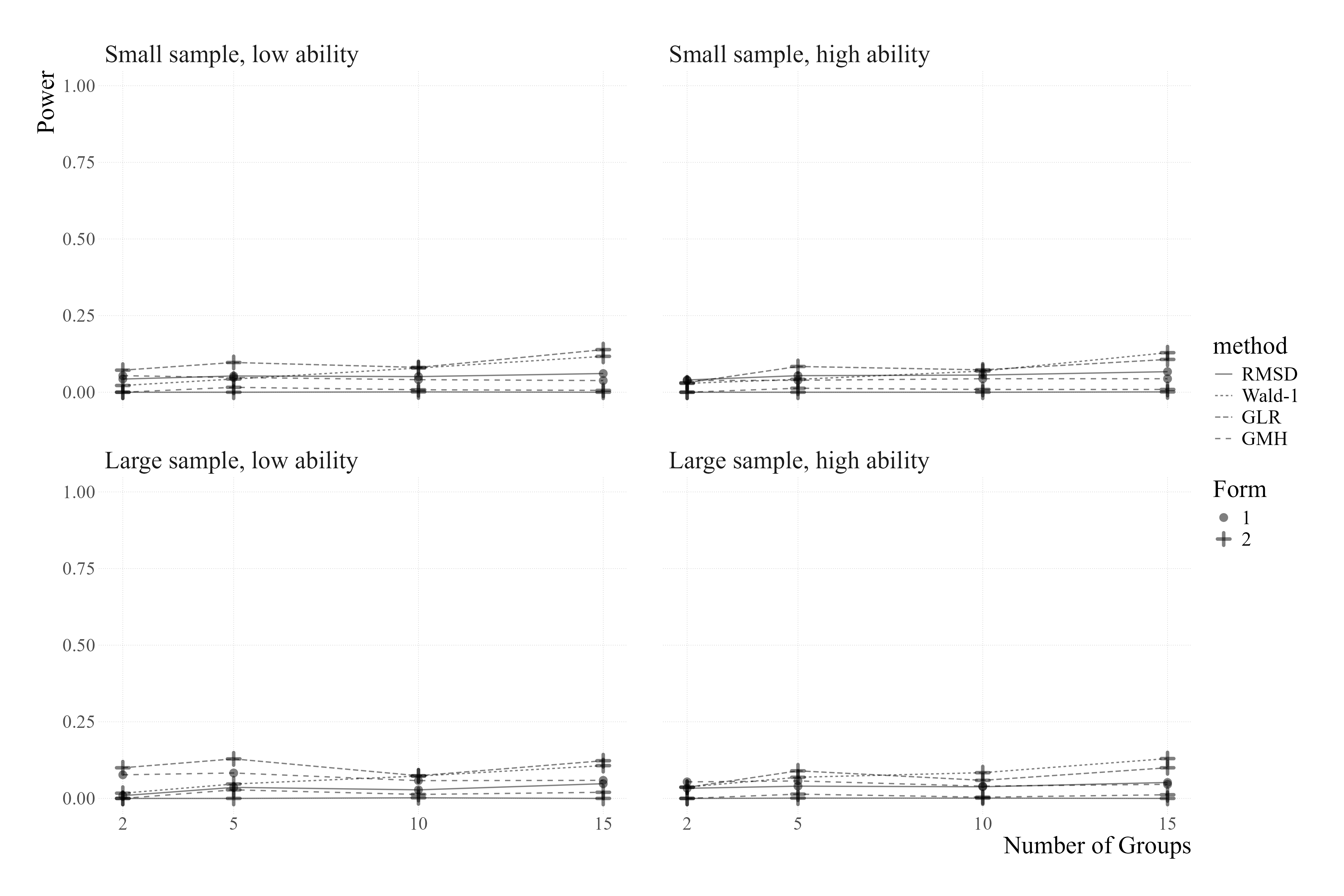}
    \caption{Powers with the selected methods in DIF detection (with DIF in $a$ for 30\% items).}
\label{fig:methods_power_4}
\end{figure}

Study 2 extends the findings from Study 1 by exploring the different DIF detection methods with DIF-contaminated conditions. Figures \ref{fig:methods_typ1_1}-\ref{fig:methods_power_4} visualize the means of Type-I and power rates for each method tested for each simulated condition. Tables \ref{tab:methods_typ1_1}-\ref{tab:methods_power_1} present the statistics for one condition (i.e., with DIF in $b$ for 20\% of items) as an example. Similar tables about other conditions are provided in supplementary materials. Since high error rates can lead to unreliable power results, we did not discuss power estimates from conditions in Study 1 that exhibited excessively high Type I error rates (i.e., when evaluating Wald-1 and GLR for nonuniform DIF detection with 15 groups).

Shown in Table \ref{tab:methods_typ1_1}, similar to what we found in Study 1, RMSD with the model-predicted cutoff had the best Type-I error rate, mostly around the middle of the acceptable range. GMH had ideal Type-I error rates when not adjusted for the $p$-value. RMSD with a fixed cutoff of 0.01 was extremely conservative, with Type-I error rates being approximately zero. Wald-1 and GLR again showed inflated Type-I error rates when detecting nonuniform DIF with 15 groups, and their Type-I error rates increased as the number of groups increased. While Type I error rates for RMSD and GMH increased slightly with the addition of DIF-contaminated items, this increase was more pronounced when DIF was present in the $b$ parameter than in the $a$ parameter.

Table \ref{tab:methods_power_1} displays powers of each method. When DIF was present in the $b$ parameter, RMSD with the model-predicted cutoff exhibited an acceptable power when analyzing five or more groups with a DIF-contaminated group featuring a small sample size and a high ability, or 15 groups with a DIF-contaminated group featuring a large sample size and a high ability. In other cases, the power to detect DIF was low. Especially, when DIF was present in the $a$ parameter, the power for all the methods remained below 0.25 across all simulated conditions. This suggests that the selected methods struggled to detect DIF associated with the $a$ paramater.

In summary, the RMSD method with the model-predicted cutoff demonstrated the best balance of controlling Type I error rates and achieving acceptable power. It was the case when DIF was associated with the $b$ parameter, the DIF-contaminated group had a high ability level, and five or more groups were involved in the analysis.

\begin{table}
\footnotesize
\def\arraystretch{0.7} 
\centering
\begin{threeparttable}
\caption{\label{tab:methods_typ1}Type-I error rates in DIF detection with the selected methods (using DIF-free data).}
\begin{tabular}{llrrrr|rrrr}
\hline
\multicolumn{1}{c}{ } & \multicolumn{1}{c}{ } & \multicolumn{4}{c}{Form 1\tnote{i}} & \multicolumn{4}{c}{Form 2\tnote{ii}} \\
\cline{3-6} \cline{7-10}
Conditions & Method & 2grps & 5grps & 10grps & 15grps & 2grps & 5grps & 10grps & 15grps\\
\hline
\multicolumn{10}{l}{\textbf{Small sample}}\\

\hspace{1em}Low ability & RMSD & 0.028 & 0.038 & 0.039 & 0.052 & 0.000 & 0.000 & 0.000 & 0.000\\

\hspace{1em} &  & (0.032) & (0.038) & (0.034) & (0.037) & (0.000) & (0.000) & (0.000) & (0.003)\\

\hspace{1em} & Wald-1 & 0.005 & 0.014 & 0.023 & 0.041 & 0.009 & 0.048 & 0.079 & 0.114\\

\hspace{1em} &  & (0.013) & (0.021) & (0.027) & (0.045) & (0.019) & (0.038) & (0.052) & (0.069)\\

\hspace{1em} & GLR & 0.023 & 0.027 & 0.025 & 0.026 & 0.031 & 0.070 & 0.072 & 0.110\\

\hspace{1em} &  & (0.028) & (0.031) & (0.032) & (0.031) & (0.033) & (0.086) & (0.058) & (0.068)\\

\hspace{1em} & GMH & 0.021 & 0.024 & 0.015 & 0.018 & 0.000 & 0.012 & 0.009 & 0.006\\

\hspace{1em} &  & (0.025) & (0.028) & (0.017) & (0.017) & (0.000) & (0.019) & (0.015) & (0.013)\\

\hspace{1em}High ability & RMSD & 0.029 & 0.032 & 0.032 & 0.044 & 0.000 & 0.000 & 0.001 & 0.001\\

\hspace{1em} &  & (0.032) & (0.032) & (0.030) & (0.033) & (0.000) & (0.003) & (0.005) & (0.006)\\

\hspace{1em} & Wald-1 & 0.004 & 0.014 & 0.025 & 0.043 & 0.013 & 0.043 & 0.072 & 0.119\\

\hspace{1em} &  & (0.011) & (0.020) & (0.027) & (0.042) & (0.019) & (0.036) & (0.050) & (0.069)\\

\hspace{1em} & GLR & 0.021 & 0.031 & 0.020 & 0.026 & 0.031 & 0.078 & 0.062 & 0.100\\

\hspace{1em} &  & (0.030) & (0.040) & (0.026) & (0.026) & (0.030) & (0.089) & (0.058) & (0.057)\\

\hspace{1em} & GMH & 0.014 & 0.018 & 0.015 & 0.019 & 0.000 & 0.009 & 0.006 & 0.005\\

\hspace{1em} &  & (0.017) & (0.017) & (0.017) & (0.017) & (0.000) & (0.015) & (0.013) & (0.012)\\

\multicolumn{10}{l}{\textbf{Large sample}}\\

\hspace{1em}Low ability & RMSD & 0.007 & 0.037 & 0.034 & 0.049 & 0.000 & 0.000 & 0.001 & 0.001\\

\hspace{1em} &  & (0.015) & (0.034) & (0.031) & (0.039) & (0.000) & (0.003) & (0.006) & (0.005)\\

\hspace{1em} & Wald-1 & 0.006 & 0.013 & 0.028 & 0.041 & 0.011 & 0.041 & 0.076 & 0.113\\

\hspace{1em} &  & (0.014) & (0.022) & (0.034) & (0.046) & (0.021) & (0.038) & (0.053) & (0.066)\\

\hspace{1em} & GLR & 0.033 & 0.028 & 0.022 & 0.031 & 0.082 & 0.076 & 0.069 & 0.107\\

\hspace{1em} &  & (0.035) & (0.036) & (0.028) & (0.037) & (0.075) & (0.087) & (0.065) & (0.072)\\

\hspace{1em} & GMH & 0.02 & 0.016 & 0.013 & 0.018 & 0.000 & 0.008 & 0.004 & 0.005\\

\hspace{1em} &  & (0.017) & (0.017) & (0.017) & (0.017) & (0.000) & (0.014) & (0.011) & (0.012)\\

\hspace{1em}High ability & RMSD & 0.024 & 0.047 & 0.042 & 0.053 & 0.000 & 0.000 & 0.001 & 0.001\\

\hspace{1em} &  & (0.028) & (0.041) & (0.035) & (0.041) & (0.000) & (0.003) & (0.006) & (0.006)\\

\hspace{1em} & Wald-1 & 0.005 & 0.019 & 0.028 & 0.047 & 0.018 & 0.042 & 0.081 & 0.117\\

\hspace{1em} &  & (0.012) & (0.027) & (0.033) & (0.049) & (0.024) & (0.032) & (0.050) & (0.068)\\

\hspace{1em} & GLR & 0.032 & 0.039 & 0.027 & 0.028 & 0.060 & 0.081 & 0.070 & 0.106\\

\hspace{1em} &  & (0.037) & (0.046) & (0.034) & (0.034) & (0.058) & (0.087) & (0.059) & (0.067)\\

\hspace{1em} & GMH & 0.019 & 0.019 & 0.017 & 0.018 & 0.000 & 0.013 & 0.007 & 0.008\\

\hspace{1em} &  & (0.017) & (0.017) & (0.017) & (0.017) & (0.000) & (0.017) & (0.014) & (0.014)\\
\hline
\end{tabular}
    \begin{tablenotes}
        \item Values in parentheses are standard deviations.
        \item[i] Using a \textit{predicted cutoff} for RMSD, \textit{uniform DIF} for Wald-1 and GLR, and \textit{non-adjusted} for GMH.
        \item[ii] Using \textit{Cutoff=0.1} for RMSD, \textit{nonuniform DIF} for Wald-1 and GLR, and \textit{adjusted} for GMH.
    \end{tablenotes}
\end{threeparttable}  
\end{table}

\begin{table}
\footnotesize
\def\arraystretch{0.7} 
\centering
\begin{threeparttable}
\caption{\label{tab:methods_typ1_1}Type-I error rates in DIF detection with the selected methods (with DIF in $b$ for 20\% items).}
\begin{tabular}{llrrrr|rrrr}
\hline
\multicolumn{1}{c}{ } & \multicolumn{1}{c}{ } & \multicolumn{4}{c}{Form 1\tnote{i}} & \multicolumn{4}{c}{Form 2\tnote{ii}} \\
\cline{3-6} \cline{7-10}
Conditions & Method & 2grps & 5grps & 10grps & 15grps & 2grps & 5grps & 10grps & 15grps\\
\hline
\multicolumn{10}{l}{\textbf{Small sample}}\\

\hspace{1em}Low ability & RMSD & 0.031 & 0.050 & 0.041 & 0.061 & 0.000 & 0.001 & 0.001 & 0.001\\

\hspace{1em} &  & (0.037) & (0.042) & (0.044) & (0.047) & (0.000) & (0.006) & (0.006) & (0.006)\\

\hspace{1em} & Wald-1 & 0.024 & 0.031 & 0.040 & 0.062 & 0.006 & 0.040 & 0.081 & 0.112\\

\hspace{1em} &  & (0.031) & (0.034) & (0.040) & (0.053) & (0.014) & (0.031) & (0.047) & (0.065)\\

\hspace{1em} & GLR & 0.018 & 0.020 & 0.021 & 0.033 & 0.022 & 0.057 & 0.067 & 0.117\\

\hspace{1em} &  & (0.028) & (0.038) & (0.036) & (0.048) & (0.037) & (0.054) & (0.048) & (0.068)\\

\hspace{1em} & GMH & 0.029 & 0.025 & 0.033 & 0.037 & 0.000 & 0.009 & 0.006 & 0.011\\

\hspace{1em} &  & (0.040) & (0.041) & (0.045) & (0.050) & (0.000) & (0.024) & (0.016) & (0.026)\\

\hspace{1em}High ability & RMSD & 0.030 & 0.057 & 0.060 & 0.070 & 0.000 & 0.001 & 0.003 & 0.002\\

\hspace{1em} &  & (0.034) & (0.046) & (0.053) & (0.052) & (0.000) & (0.006) & (0.012) & (0.010)\\

\hspace{1em} & Wald-1 & 0.031 & 0.040 & 0.045 & 0.062 & 0.011 & 0.037 & 0.076 & 0.114\\

\hspace{1em} &  & (0.035) & (0.042) & (0.038) & (0.052) & (0.017) & (0.035) & (0.052) & (0.060)\\

\hspace{1em} & GLR & 0.011 & 0.017 & 0.023 & 0.040 & 0.019 & 0.052 & 0.070 & 0.107\\

\hspace{1em} &  & (0.029) & (0.032) & (0.035) & (0.049) & (0.024) & (0.053) & (0.056) & (0.066)\\

\hspace{1em} & GMH & 0.022 & 0.028 & 0.031 & 0.043 & 0.000 & 0.010 & 0.011 & 0.009\\

\hspace{1em} &  & (0.042) & (0.045) & (0.047) & (0.050) & (0.000) & (0.019) & (0.022) & (0.021)\\

\multicolumn{10}{l}{\textbf{Large sample}}\\

\hspace{1em}Low ability & RMSD & 0.008 & 0.047 & 0.035 & 0.044 & 0.000 & 0.000 & 0.001 & 0.000\\

\hspace{1em} &  & (0.018) & (0.041) & (0.035) & (0.037) & (0.000) & (0.004) & (0.006) & (0.004)\\

\hspace{1em} & Wald-1 & 0.022 & 0.030 & 0.054 & 0.067 & 0.016 & 0.038 & 0.085 & 0.116\\

\hspace{1em} &  & (0.029) & (0.031) & (0.041) & (0.055) & (0.019) & (0.039) & (0.053) & (0.054)\\

\hspace{1em} & GLR & 0.015 & 0.032 & 0.017 & 0.028 & 0.042 & 0.103 & 0.073 & 0.108\\

\hspace{1em} &  & (0.029) & (0.048) & (0.029) & (0.045) & (0.061) & (0.097) & (0.056) & (0.062)\\

\hspace{1em} & GMH & 0.017 & 0.033 & 0.029 & 0.030 & 0.000 & 0.008 & 0.007 & 0.006\\

\hspace{1em} &  & (0.032) & (0.043) & (0.048) & (0.042) & (0.000) & (0.020) & (0.017) & (0.019)\\

\hspace{1em}High ability & RMSD & 0.038 & 0.052 & 0.037 & 0.044 & 0.000 & 0.000 & 0.001 & 0.001\\

\hspace{1em} &  & (0.041) & (0.048) & (0.037) & (0.037) & (0.000) & (0.004) & (0.007) & (0.006)\\

\hspace{1em} & Wald-1 & 0.047 & 0.056 & 0.071 & 0.083 & 0.018 & 0.038 & 0.088 & 0.112\\

\hspace{1em} &  & (0.054) & (0.047) & (0.050) & (0.053) & (0.021) & (0.036) & (0.050) & (0.056)\\

\hspace{1em} & GLR & 0.023 & 0.034 & 0.023 & 0.050 & 0.029 & 0.081 & 0.067 & 0.098\\

\hspace{1em} &  & (0.081) & (0.049) & (0.032) & (0.060) & (0.033) & (0.080) & (0.037) & (0.063)\\

\hspace{1em} & GMH & 0.034 & 0.050 & 0.045 & 0.064 & 0.000 & 0.014 & 0.016 & 0.017\\

\hspace{1em} &  & (0.062) & (0.071) & (0.066) & (0.071) & (0.000) & (0.025) & (0.031) & (0.034)\\
\hline
\end{tabular}
    \begin{tablenotes} 
        \item Values in parentheses are standard deviations.
        \item[i] Using a \textit{predicted cutoff} for RMSD, \textit{uniform DIF} for Wald-1 and GLR, and \textit{non-adjusted} for GMH.
        \item[ii] Using \textit{Cutoff=0.1} for RMSD, \textit{nonuniform DIF} for Wald-1 and GLR, and \textit{adjusted} for GMH.
    \end{tablenotes}
\end{threeparttable}  
\end{table}

\begin{table}
\footnotesize
\def\arraystretch{0.7} 
\centering
\begin{threeparttable}
\caption{\label{tab:methods_power_1}Powers in DIF detection with the selected methods (with DIF in $b$ for 20\% items).}
\begin{tabular}{llrrrr|rrrr}
\hline
\multicolumn{1}{c}{ } & \multicolumn{1}{c}{ } & \multicolumn{4}{c}{Form 1\tnote{i}} & \multicolumn{4}{c}{Form 2\tnote{ii}} \\
\cline{3-6} \cline{7-10}
Conditions & Method & 2grps & 5grps & 10grps & 15grps & 2grps & 5grps & 10grps & 15grps\\
\hline
\multicolumn{10}{l}{\textbf{Small sample}}\\

\hspace{1em}Low ability & RMSD & 0.240 & 0.368 & 0.330 & 0.373 & 0.000 & 0.038 & 0.072 & 0.100\\

\hspace{1em} &  & (0.168) & (0.163) & (0.177) & (0.176) & (0.000) & (0.070) & (0.093) & (0.109)\\

\hspace{1em} & Wald-1 & 0.027 & 0.037 & 0.052 & 0.060 &  &  &  & \\

\hspace{1em} &  & (0.061) & (0.073) & (0.088) & (0.090) &  &  &  & \\

\hspace{1em} & GLR & 0.187 & 0.140 & 0.117 & 0.122 &  &  &  & \\

\hspace{1em} &  & (0.148) & (0.144) & (0.133) & (0.144) &  &  &  & \\

\hspace{1em} & GMH & 0.242 & 0.185 & 0.148 & 0.135 & 0.000 & 0.127 & 0.073 & 0.070\\

\hspace{1em} &  & (0.175) & (0.179) & (0.176) & (0.151) & (0.000) & (0.123) & (0.112) & (0.114)\\

\hspace{1em}High ability & RMSD & 0.343 & 0.630 & 0.597 & 0.660 & 0.002 & 0.190 & 0.247 & 0.280\\

\hspace{1em} &  & (0.195) & (0.202) & (0.190) & (0.174) & (0.017) & (0.128) & (0.147) & (0.160)\\

\hspace{1em} & Wald-1 & 0.053 & 0.078 & 0.068 & 0.070 &  &  &  & \\

\hspace{1em} &  & (0.097) & (0.115) & (0.111) & (0.106) &  &  &  & \\

\hspace{1em} & GLR & 0.248 & 0.243 & 0.207 & 0.152 &  &  &  & \\

\hspace{1em} &  & (0.141) & (0.170) & (0.161) & (0.157) &  &  &  & \\

\hspace{1em} & GMH & 0.298 & 0.312 & 0.257 & 0.195 & 0.000 & 0.252 & 0.163 & 0.103\\

\hspace{1em} &  & (0.197) & (0.236) & (0.224) & (0.188) & (0.000) & (0.173) & (0.150) & (0.116)\\

\multicolumn{10}{l}{\textbf{Large sample}}\\

\hspace{1em}Low ability & RMSD & 0.068 & 0.098 & 0.108 & 0.150 & 0.000 & 0.000 & 0.010 & 0.007\\

\hspace{1em} &  & (0.092) & (0.116) & (0.131) & (0.139) & (0.000) & (0.000) & (0.040) & (0.033)\\

\hspace{1em} & Wald-1 & 0.03 & 0.047 & 0.052 & 0.073 &  &  &  & \\

\hspace{1em} &  & (0.069) & (0.089) & (0.097) & (0.109) &  &  &  & \\

\hspace{1em} & GLR & 0.162 & 0.177 & 0.157 & 0.168 &  &  &  & \\

\hspace{1em} &  & (0.145) & (0.129) & (0.148) & (0.143) &  &  &  & \\

\hspace{1em} & GMH & 0.208 & 0.208 & 0.185 & 0.197 & 0.000 & 0.157 & 0.132 & 0.117\\

\hspace{1em} &  & (0.183) & (0.178) & (0.177) & (0.176) & (0.000) & (0.123) & (0.126) & (0.129)\\

\hspace{1em}High ability & RMSD & 0.462 & 0.393 & 0.485 & 0.610 & 0.022 & 0.018 & 0.070 & 0.177\\

\hspace{1em} &  & (0.195) & (0.186) & (0.177) & (0.168) & (0.061) & (0.052) & (0.095) & (0.151)\\

\hspace{1em} & Wald-1 & 0.078 & 0.087 & 0.085 & 0.092 &  &  &  & \\

\hspace{1em} &  & (0.143) & (0.137) & (0.122) & (0.115) &  &  &  & \\

\hspace{1em} & GLR & 0.312 & 0.362 & 0.312 & 0.285 &  &  &  & \\

\hspace{1em} &  & (0.218) & (0.193) & (0.183) & (0.178) &  &  &  & \\

\hspace{1em} & GMH & 0.368 & 0.435 & 0.395 & 0.373 & 0.000 & 0.342 & 0.323 & 0.288\\

\hspace{1em} &  & (0.261) & (0.269) & (0.283) & (0.272) & (0.000) & (0.215) & (0.192) & (0.189)\\
\hline
\end{tabular}
    \begin{tablenotes}
        \item Values in parentheses are standard deviations.
        \item[i] Using a \textit{predicted cutoff} for RMSD, \textit{uniform DIF} for Wald-1 and GLR, and \textit{non-adjusted} for GMH.
        \item[ii] Using \textit{Cutoff=0.1} for RMSD, \textit{nonuniform DIF} for Wald-1 and GLR, and \textit{adjusted} for GMH.
    \end{tablenotes}
\end{threeparttable}  
\end{table}

\section{Discussion}

Our study contributed to the advancement of the current knowledge base. Firstly, our results showed that the RMSD approach used in LSAs should use a model-predicted cutoff instead of a fixed cutoff like 0.1. We found that using a fixed cutoff made RMSD overly conservative, confirming the finding in \textcite{tijmstra_sensitivity_2020} that no fixed cutoff could universally fit the RMSD approach. Secondly, we filled the literature gap about GMH's performance hwne using the $p$-value adjustment. GMH was found to exhibit an acceptable Type-I error rate without a $p$-value adjustment. Finally, this study provided the first simulation-based evaluation of GLR’s performance and compared it with other multi-group methods.
\\
Our simulation design aligned with \textcite{penfield_assessing_2001} in that we consistently set the focal groups to have higher parameter estimates. However, unlike their study, we set only one focal group to be DIF-contaminated to ensure simplicity and interpretability. The low power of GMH found in this study confirmed the low-power situation in \textcite{penfield_assessing_2001} for GMH with a group sample size smaller than 500 when the $b$ parameter is DIF-contaminated. While we revealed that all the four tested methods exhibited low power to detect DIF in the $a$ parameter, this pattern aligned with what was found in \textcite{lim_residualbased_2022} when assessing powers for nonuniform DIF detection. Our reported low power of Wald-1 partly contradicted the high power (over 90\%) reported for Wald-1 in \textcite{woods_langer-improved_2013}. However, their simulation design differed from ours, as they set all the focal groups to be DIF-contaminated and introduced DIF in opposite directions for the $a$ and $b$ parameters in the focal groups.
\\
This study shed light on future research directions for multi-group DIF methods. First, exploring how the sample size and the number of DIF groups interact in DIF studies would be valuable. \textcite{penfield_assessing_2001} found GMH power increased significantly with a sample size of 500 or more per group. Past studies have manipulated the number of DIF-contaminated focal groups (e.g., 2-10 or more) when holding the sample size constant for each group (e.g., 500, 1000, 1500 per group or in cherry-picked combinations). In contrast, our study adopted a more ecologically valid approach by utilizing the real-world sample sizes of education systems participating in the TIMSS 2019 as the sample sizes for each group. These real-world sample sizes varied considerably across groups, ranging from 326 to 1480. While this approach aligned the analysis with the complexities of actual testing situations, it presented a challenge for future research, as the power of DIF detection methods might be impacted by uneven sample sizes across groups. Future studies could explore methods that are more robust to such variations in sample size.
\\
Second, future studies could systematically vary the direction of DIF in the focal group relative to the reference group (e.g., opposite vs. same direction for $a$ and $b$ parameters). \textcite{woods_langer-improved_2013} set opposite directions for DIF in the $a$ and $b$ parameters, whereas \textcite{penfield_assessing_2001} used the same direction. Further exploration is needed in this regard to check the associated impact on DIF results. Third, how the magnitude of DIF would vary given the difference in the $a$ and $b$ parameters would be worth investigating. The same amount of shift in parameter estimates may vary in impact given the parameter (e.g., 0.4 might be large for $a$ but not $b$). Importantly, further research would be needed to develop methods more robust to DIF in the $a$ parameter, as prior research on DIF detection mostly focused on methods' performance in identifying DIF in the $b$ parameter about item difficulty.
\\
In addition, research is needed in fundamental methodological aspects of multi-group DIF analysis. First, the effect size of DIF in multi-group DIF analysis has not been extensively studied. Currently, no method provides estimates of the effect size in multi-group DIF analysis \parencite{penfield_assessing_2001}. Second, the recovery of asymptotic covariance matrices remains a challenge for IRT-based DIF methods. This issue has been a long-standing problem when estimating IRT models with the EM algorithm \parencite{cai_sem_2008}. 
Third, more evidence is needed to investigate possible benefits of the $p$-value adjustment for Wald-1. Our findings showed that the Type-I error rate of Wald-1 increased as the number of groups increased, warranting further investigation of solutions such as the $p$-value adjustment in this regard.

\section{Conclusion}

We compared four frequentist, unidimensional methods in terms of their Type-I error and power via Monte Carlo simulation of controlled assessment settings. Our findings provided valuable insights into the strengths and limitations of the different DIF detection methods under various conditions. These insights are crucial for selecting appropriate methods based on the specific requirements and conditions of DIF analysis. Our primary finding is that RMSD with model-predicted cutoffs (i.e., 0.060, 0.070, 0.075, and 0.075 for 2-, 5-, 10-, and 15-group analyses, respectively) yielded the best Type-I error rates. While power rates were low across the methods, RMSD with model-predicted cutoffs exhibited a relatively acceptable power when analyzing five or more groups with a DIF-contaminated group featuring a small sample size and a high ability, or 15 groups with a DIF-contaminated group featuring a large sample size and a high ability. The fixed cutoff of 0.01 used in TIMSS is not suitable for RMSD, as it results in overly low Type-I error rates and powers. Caution should be exercised when using any of the tested methods here when expecting DIF to occur in the $a$ parameter of an item, as these methods are not sensitive enough to DIF associated with this parameter. 

\newpage

\printbibliography

\newpage

\section{Statements and Declarations}

\subsection{Disclosure Statement}
The authors report there are no competing interests to declare.

\subsection{Acknowledgements}
The authors want to express sincere appreciation and gratitude to many people who provided helpful feedback on early drafts of this manuscript, including Dr. Jane Rogers at the University of Connecticut, Drs. Carolyn Anderson and Justin Kern at the University of Illinois Urbana-Champaign, and the journal editor and anonymous reviewers. Also, the authors appreciate the support and opportunities to present preliminary findings at national conferences and weekly seminars of the Quantitative and Qualitative Methodology, Measurement, and Evaluation (QUERIES) program at the College of Education, University of Illinois Urbana-Champaign. This work was supported by the Conference Travel Grant at the College of Education, University of Illinois Urbana-Champaign.

\end{document}